\documentclass[sigconf]{acmart}
\usepackage{multirow} 
\usepackage{enumitem}
\usepackage{makecell}

\AtBeginDocument{%
  }

\copyrightyear{2024}
\acmYear{2024}
\setcopyright{acmlicensed}\acmConference[CIKM '24]{Proceedings of the 33rd ACM International Conference on Information and Knowledge Management}{October 21--25, 2024}{Boise, ID, USA}
\acmBooktitle{Proceedings of the 33rd ACM International Conference on Information and Knowledge Management (CIKM '24), October 21--25, 2024, Boise, ID, USA}
\acmDOI{10.1145/3627673.3679655}
\acmISBN{979-8-4007-0436-9/24/10}

\begin{document}

\title{Context Matters: Enhancing Sequential Recommendation with Context-aware Diffusion-based Contrastive Learning}
\author{Ziqiang Cui}
\affiliation{
  \institution{City University of Hong Kong}
  \country{Hong Kong SAR}
}
\email{ziqiang.cui@my.cityu.edu.hk}

\author{Haolun Wu}
\affiliation{
  \institution{McGill University}
  \city{Montréal}
  \country{Canada}
}
\email{haolun.wu@mail.mcgill.ca}

\author{Bowei He}
\affiliation{
  \institution{City University of Hong Kong}
 \country{Hong Kong SAR}
}
\email{boweihe2-c@my.cityu.edu.hk}

\author{Ji Cheng}
\affiliation{
  \institution{City University of Hong Kong}
 \country{Hong Kong SAR}
}
\email{J.Cheng@my.cityu.edu.hk}

\author{Chen Ma}
\authornote{Corresponding author.}
\affiliation{
  \institution{City University of Hong Kong}
  \country{Hong Kong SAR}
}
\email{chenma@cityu.edu.hk}

\begin{abstract} 
Contrastive learning has been effectively utilized to enhance the training of sequential recommendation models by leveraging informative self-supervised signals. Most existing approaches generate augmented views of the same user sequence through random augmentation and subsequently maximize their agreement in the representation space. However, these methods often neglect the rationality of the augmented samples. Due to significant uncertainty, random augmentation can disrupt the semantic information and interest evolution patterns inherent in the original user sequences. Moreover, pulling semantically inconsistent sequences closer in the representation space can render the user sequence embeddings insensitive to variations in user preferences, which contradicts the primary objective of sequential recommendation.
To address these limitations, we propose the \textbf{C}ontext-\textbf{a}ware \textbf{Di}ffusion-based Contrastive Learning for Sequential \textbf{Rec}ommendation, named \textbf{CaDiRec}. The core idea is to leverage context information to generate more reasonable augmented views. Specifically, CaDiRec employs a context-aware diffusion model to generate alternative items for the given positions within a sequence. These generated items are aligned with their respective context information and can effectively replace the corresponding original items, thereby generating a positive view of the original sequence. By considering two different augmentations of the same user sequence, we can construct a pair of positive samples for contrastive learning.
To ensure representation cohesion, we train the entire framework in an end-to-end manner, with shared item embeddings between the diffusion model and the recommendation model. Extensive experiments on five benchmark datasets demonstrate the advantages of our proposed method over existing baselines.
The code of our method is available at \textcolor{blue}{\url{https://github.com/ziqiangcui/CaDiRec}}.
\end{abstract}



\maketitle

\section{Introduction}
Sequential recommendation (SR) systems predict the next item for users 
based on their historical interactions, 
which have demonstrated significant value on various 
online platforms.
One of the major challenges in SR is data sparsity \cite{qin2023meta,zhou2023equivariant}. The limited and noisy user interaction records impede the training of complex SR models, thereby constraining their performance. 
Recently, contrastive learning has emerged as a promising approach to address this challenge. By directly extracting inherent data correlations from user sequences, contrastive learning enhances user representation learning, resulting in notable advancements in the field \cite{chen2022intent,qiu2022contrastive,xie2022contrastive}.

Existing methods typically use data augmentation to create augmented views of original user sequences and maximize the agreement among different views of the same user.
In terms of data augmentation levels, existing methods can be categorized into three types: 1) \textit{Data Level}. 
This involves generating augmented views of user sequences by applying random augmentations \cite{xie2022contrastive} such as masking, substituting, reordering, and cropping. More informative methods based on item correlation are also used \cite{liu2021contrastive}.
2) \textit{Model Level}. To reduce the disturbance to original sequences, some methods propose model-level operations \cite{qiu2022contrastive,liu2022improving}. These involve performing a forward pass of neural networks on each user sequence twice, each time with a different dropout mask. However, the dropout operation still introduces a considerable amount of uncertainty due to its randomness.
3) \textit{Mixed Level}. 
These approaches, exemplified by Qin et al. \cite{qin2023meta} and Zhou et al. \cite{zhou2023equivariant}, integrate both data-level and model-level augmentations to extract more expressive features and establish distinct contrastive objectives for varying levels of augmentation.


\begin{figure}[t]
\setlength{\belowcaptionskip}{-5mm} 
  \centering
  \includegraphics[width=0.75\columnwidth]{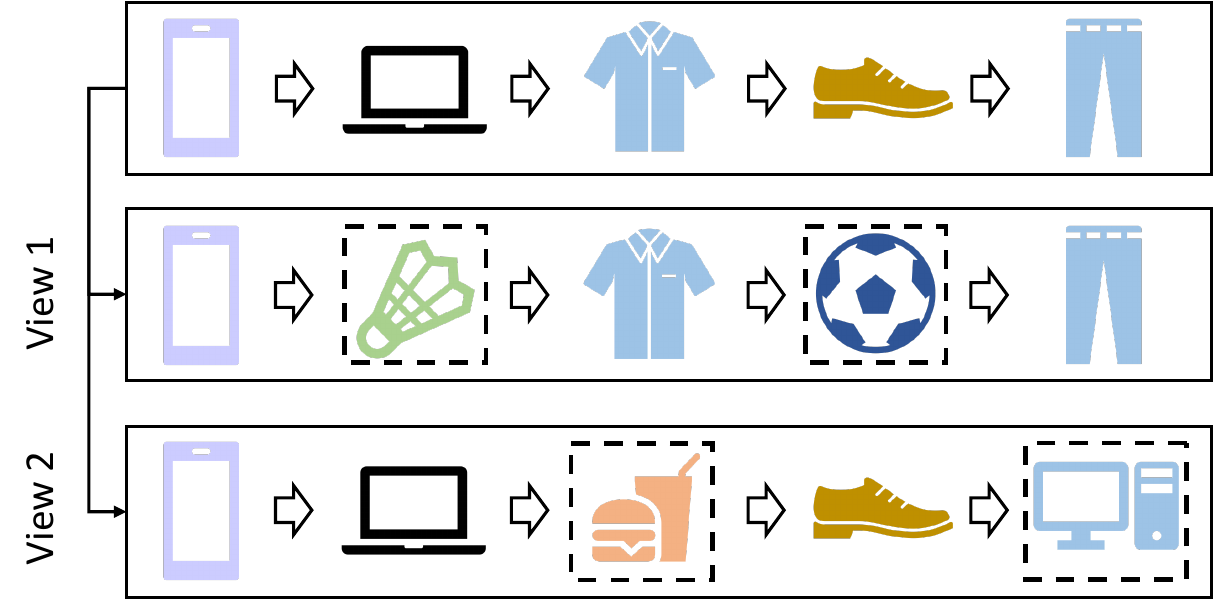}
  \caption{
  An example of augmented sequences with semantic discrepancies, where view 1 and view 2 are two augmented views of the original user sequence by random substitution.} 
   \label{intro_fig}
\end{figure}

Despite the demonstrated effectiveness of the aforementioned studies, they neglect the rationality of data augmentation. Most of them \cite{xie2022contrastive, chen2022intent, qin2023meta, qiu2022contrastive,liu2022improving,zhou2023equivariant} employ random augmentation, either at the data or model level, to generate augmented views. However, this approach introduces a considerable amount of uncertainty, leading to significant semantic discrepancies between the pairs of augmented views.
For example, Figure \ref{intro_fig} shows an original user sequence and two augmented views obtained by applying the random substitution operation twice. It is evident that these two augmented sequences exhibit substantial semantic discrepancies. The preference of View 1 focuses on sports and clothing, while View 2 mainly concentrates on electronic products. In addition, the evolution of user interests in the two views also exhibits distinctly different patterns. Maximizing the representation agreement between such augmented views can result in user representations becoming insensitive to varying user preferences and interest evolution patterns, which contradicts the primary objective of SR.

How can we generate more reasonable augmented views for better contrastive learning? Intuitively, effective data augmentation should consider the unique characteristics of SR. SR differs from other applications, such as computer vision, in two main aspects: 1) interaction records are often sparse, making user sequences highly sensitive to modifications, and 2) there is a strong sequential interdependence between items in the sequence. Therefore, when modifying items within a user sequence, failing to consider the preceding and subsequent items and their sequential dependencies can lead to a complete change in the sequence's semantics. In light of this, we make the first attempt to introduce context information to improve the rationality of augmented views. 
Our basic idea involves learning the conditional distribution of each position within a sequence based on its contextual information. By generating replacements according to this conditional distribution, we can produce augmented sequences.

To achieve this goal, we propose the \textbf{C}ontext-\textbf{a}ware \textbf{{Di}}ffusion-based Contrastive Learning for Sequential 
\textbf{{Rec}}ommendation (\textit{\textbf{CaDiRec}}). Specifically, we employ a diffusion model as the sample generator due to its remarkable capabilities in learning underlying data distributions and robust conditional generation \cite{ho2020denoising, li2022diffusion}. To encode the context information and capture the complex sequential dependencies, we adopt the bidirectional Transformer \cite{devlin2018bert} as the encoder for the diffusion model. The encoded context representation guides the diffusion model in gradually refining the generated items, thereby enabling it to accurately learn the conditional distribution of items within a sequence. During contrastive learning, CaDiRec generates alternative items by sampling from the learned conditional distribution. This ensures that the generated items are coherent with the context and sequential dependencies, thereby producing more reasonable augmented sequences.
Moreover, to align the embedding space of the diffusion model with that of the SR model, we train both models jointly with shared item embeddings in an end-to-end manner. By integrating these designs, CaDiRec effectively enhances the quality of data augmentation, leading to better contrastive learning and improved user modeling.

Our main contributions are summarized as follows:
\begin{itemize}[leftmargin=1em]
\item We propose a novel model, CaDiRec, that generates reasonable augmented views for contrastive learning through conditional generation, thereby improving sequential recommendation.
\item To the best of our knowledge, this is the first work to explore the use of context information (i.e., both preceding and succeeding items) for contrastive learning in sequential recommendation.
\item We conduct extensive experiments on five public benchmark
datasets, and the results demonstrate the superiority of our method.
\end{itemize}


\section{Related Work}
In this section, we summarize the related works from the following three fields: \textit{(\romannumeral1)} sequential recommendation, \textit{(\romannumeral2)} contrastive learning, and \textit{(\romannumeral3)} diffusion models. 
\subsection{Sequential Recommendation}
In the initial phase, researchers treated the evolution of user behaviors as a Markov process~\cite{rendle2010factorizing, he2016fusing}.
With the rapid advancements in deep learning, various techniques such as convolutional neural networks (CNN) and recurrent neural networks (RNN) have been utilized~\cite{tang2018personalized,hidasi2015session,hidasi2018recurrent}, leading to remarkable achievements. 
Subsequently, the introduction of the attention mechanism has significantly improved the SR performance. 
SASRec~\cite{kang2018self}, for instance, is the pioneering work that applied the self-attention mechanism to SR. 
Following that, BERT4Rec~\cite{sun2019bert4rec} is proposed to use a bidirectional self-attention encoder to capture context information of the user sequence. 
In recent years, many self-attention-based and graph-based methods have made improvements to existing approaches, achieving notable progress \cite{he2020lightgcn,he2023dynamically,zhang2022dynamic,he2023dynamic,cui2022dual,zhang2022enhanced,xia2022multi}.

\subsection{Contrastive Learning}
Contrastive learning, which enhances representation learning by constructing informative supervisory signals from unlabeled data, has been extensively applied in various domains such as computer vision (CV) \cite{li2020prototypical, chen2020simple, chen2021exploring} and natural language processing (NLP) \cite{yan2021consert}. Given the inherent issues of user behavior sparsity and noisy interaction records in recommendation scenarios, contrastive learning has also played a crucial role in multiple recommendation tasks \cite{zhang2024recdcl,jiang2023diffkg, yu2022graph, wu2021self, xia2023automated, yin2022autogcl, yang2023debiased}.
When it comes to SR, $\rm {S^3\text{-}Rec}$~\cite{zhou2020s3} first introduces a self-supervised method that incorporates auxiliary objectives to learn the correlations among items, attributes, and segments. 
CL4SRec~\cite{xie2022contrastive} designs three data-level augmentation operators, namely crop, mask, and reorder, to generate positive pairs and promote the invariance of their representations. 
However, introducing random perturbations to the already sparse interaction records of a user can alter her original preference. CoSeRec~\cite{liu2021contrastive} suggests substituting a specific item in the sequence with a similar item. 
However, the item similarity is measured by simple co-occurrence counts or item embedding distance, neglecting the context information of user behaviors. DuoRec~\cite{qiu2022contrastive} proposes a model-level augment strategy, which generates positive pairs by forward-passing an input sequence twice with different dropout masks. However, this approach also introduces a significant amount of randomness, lacking the ability to maintain semantic consistency.
In addition, ICLRec~\cite{chen2022intent} extracts user intent from sequential information and performs contrastive learning between user representations and intent representations. 
ECL-SR~\cite{zhou2023equivariant} designs multiple contrastive objectives for augmented views at different levels.
MCLRec~\cite{qin2023meta} further combines data-level and model-level augmentation strategies, which applies random data augmentation proposed by CL4SRec to the input sequence and then feed the augmented data into MLP layers for the model-level augment. 
Despite the effectiveness of the aforementioned
studies, the design intentions of these methods do not reflect the constraints on semantic consistency in the augmented views, which can potentially lead to the generation of unreasonable samples for contrastive learning.

\subsection{Diffusion Models}
Diffusion Models have demonstrate superior generative capabilities compared to alternative methods such as GANs~\cite{goodfellow2014generative} and VAEs~\cite{kingma2013auto} in diverse generative tasks, such as image synthesis~\cite{ho2020denoising,song2020score,dhariwal2021diffusion} and text generation~\cite{li2022diffusion,gong2022diffuseq}, which can be attributed to their precise approximation of the underlying data generation distribution and provision of enhanced training stability.
Recently, diffusion models have also been employed in the field of SR. Some methods~\cite{yang2023generate,wang2023conditional,wang2023diffusion,li2023diffurec,du2023sequential} directly utilize diffusion models as the fundamental architecture for SR. In contrast, other methods~\cite{liu2023diffusion,wu2023diff4rec} adopt a two-stage paradigm. They train a diffusion model to generate pseudo user interactions aimed at expanding the original user sequences. These augmented datasets are then used to train downstream recommendation models. It should be noted that they only leverage the unidirectional information of user sequences as the guidance during the diffusion process. 
In contrast, we utilize both left and right context information as guidance, generating semantic-consistent augmented samples for contrastive learning in an end-to-end manner. To the best of our knowledge, this is the first instance of employing diffusion models for contrastive learning in SR.

\section{Preliminary}
In this section, we first define our problem statement, followed by introducing basic principles of diffusion models.
\subsection{Problem Statement}
 The primary objective of sequential recommendation is to provide personalized recommendations for the next item to users, leveraging their historical interactions. We denote the user and item sets as $\mathcal{U}$ and $\mathcal{V}$, respectively. Each user $u\in \mathcal{U}$ has a chronological sequence of interacted items $\mathbf{s}^u = [v^u_1, v^u_2..., v^u_{|\mathbf{s}^u|}]$, where $v^u_t$ indicates the item that $u$ interacted with at step $t$, and $|\mathbf{s}^u|$ is the number of interacted items of user $u$. The goal is to predict the next item at time step $|\mathbf{s}^u|+1$ according to $\mathbf{s}^u$, which can be formulated as:
 \begin{equation}
     \mathop{\arg\max}\limits_{v_i \in \mathcal{V}}  {P(v_{|\mathbf{s}^u|+1}=v_i|\mathbf{s}^u)},
 \end{equation}
 where the probability $P$ represents the likelihood of item $v_i$ being the next item, conditioned on $\mathbf{s}^u$. 
\subsection{Diffusion Models}
We provide an introduction to the fundamental principles of diffusion models based on DDPM~\cite{ho2020denoising}. Typically, a diffusion model consists of forward and reverse processes. Given a data point sampled from a real-world data distribution $\mathbf{x}_0 \sim q(\mathbf{x}_0)$, the forward process gradually corrupts $\mathbf{x}_0$ into a standard Gaussian noise $\mathbf{x}_T \sim  N(0; \mathbf{I})$, which is formulated as:
\begin{equation}
\label{eq2}
\begin{aligned}
    q(\mathbf{x}_{1:T}|\mathbf{x}_0)&= \prod_{t=1}^{T} q(\mathbf{x}_t|\mathbf{x}_{t-1}),\\
    q(\mathbf{x}_t|\mathbf{x}_{t-1})&= \mathcal{N}(\mathbf{x}_t;\sqrt{1-\beta_t}\mathbf{x}_{t-1},\beta_t \mathbf{I}),
\end{aligned}
\end{equation}
where $\beta_t \in (0,1)$ is the variance scale at time step $t$.

After the completion of the forward process, the reverse denoising process aims to gradually reconstruct the original data $\mathbf{x}_0$. This is achieved by sampling from $\mathbf{x}_T$ using a learned diffusion model, which can be formulated as:
\begin{equation}
    \begin{aligned}
p_{\theta}(\mathbf{x}_{0:T}) &=p(\mathbf{x}_{T}) \prod_{t=1}^T p_\theta(\mathbf{x}_{t-1}|\mathbf{x}_t), \\ 
p_\theta(\mathbf{x}_{t-1}|\mathbf{x}_t) &= \mathcal{N} \big(\mathbf{x}_{t-1}; \mu_\theta (\mathbf{x}_t, t),\Sigma_\theta(\mathbf{x}_t,t) \big).
    \end{aligned}
\end{equation} 
Training can be performed by optimizing the variational lower bound of $\log p_\theta(\mathbf{x}_0)$:
\begin{align}
\label{vlb_loss}
    \mathcal{L}_\text{vlb}(\mathbf{x}_0) = \mathop{\mathbb E} \limits_{q(\mathbf{x}_{1:T}|\mathbf{x}_0)} 
    [ &\log \frac{q(\mathbf{x}_T|\mathbf{x}_0)}{p_\theta (\mathbf{x}_T)} \notag\\
    &+ \sum_{t=2}^T \log \frac{q(\mathbf{x}_{t-1}|\mathbf{x}_0,\mathbf{x}_t)}{p_\theta (\mathbf{x}_{t-1}|\mathbf{x}_t)}-\log p_\theta(\mathbf{x}_0|\mathbf{x}_1) ].
\end{align}
\citet{ho2020denoising} further propose to utilize the KL divergence for more efficient training, which directly compares $p_{\theta}(\mathbf{x}_{t-1}|\mathbf{x}_t)$ against forward process posteriors, resulting in a mean-squared error loss:
\begin{equation}
 \mathcal{L}_\text{simple}(\mathbf{x}_0) = \sum_{t=1}^T \mathop{\mathbb E} \limits_{q(\mathbf{x}_t | \mathbf{x}_0)} 
 ||\mu_\theta(\mathbf{x}_t, t) - \hat{\mu}(\mathbf{x}_t,\mathbf{x}_0) || ^2 ,
\end{equation}
where $\mu_\theta(\mathbf{x}_t, t)$ is the predicted mean of $p_\theta(\mathbf{x}_{t-1} | \mathbf{x}_t)$ computed by a neural network, and $\hat{\mu}(\mathbf{x}_t,\mathbf{x}_0)$ is the mean of the posterior $ q(\mathbf{x}_{t-1} | \mathbf{x}_t,\mathbf{x}_0)$, which is tractable when conditioned on $\mathbf{x}_0$.




\section{Methodology}
\begin{figure*}[t]
  \centering
  \includegraphics[width=0.95\textwidth]{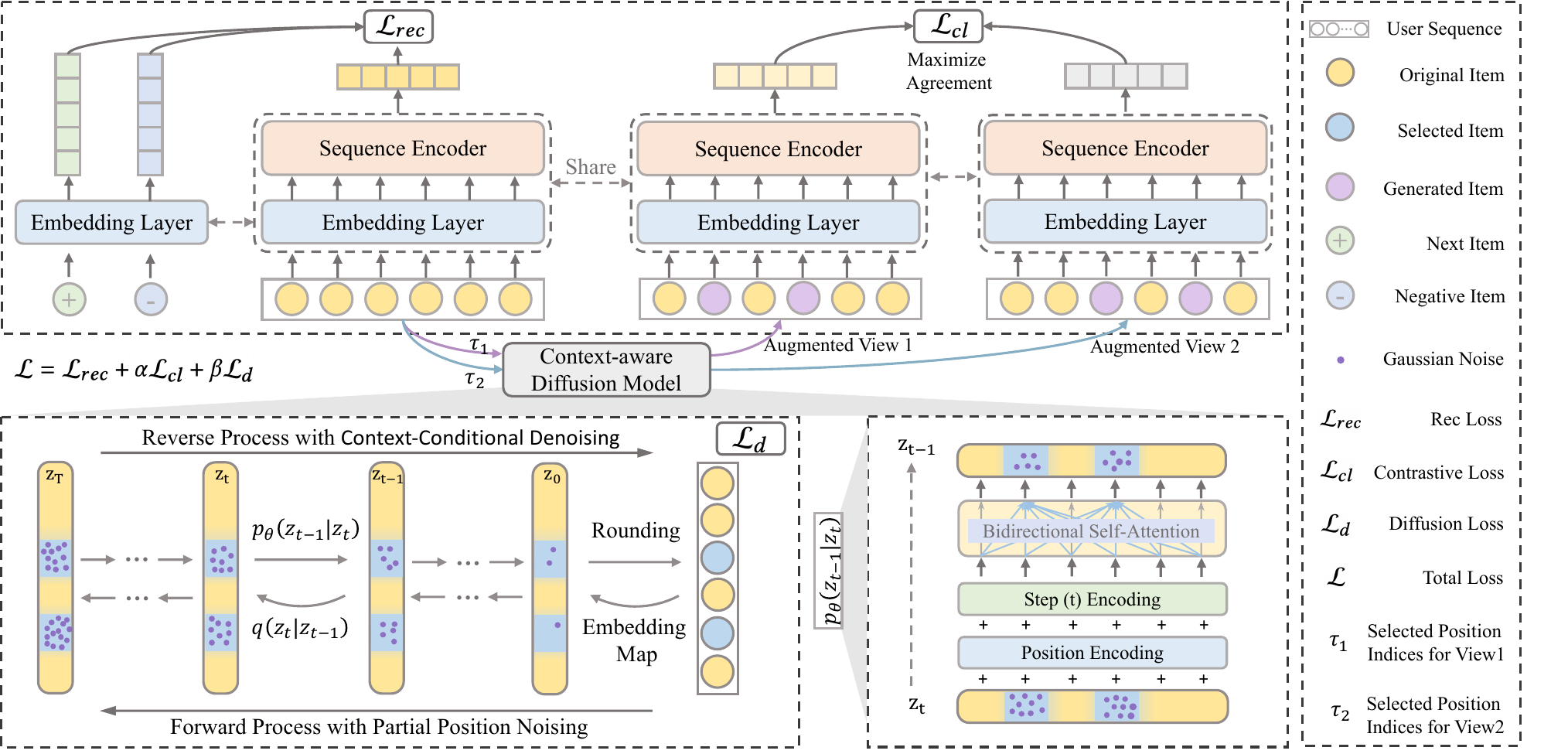}
  \caption{Overview of our proposed CaDiRec. CaDiRec employs a context-aware diffusion model to generate reasonable augmented views for contrastive learning. The context-aware diffusion model comprises a forward process with partial position noising and a reverse process with context-conditional denoising. These designs enable the model to generate contextually appropriate substitutions for selected positions, leading to the production of reasonable augmented sequences. To effectively capture contextual dependencies, CaDiRec employs a bidirectional Transformer architecture within the diffusion model.
  } 
   \label{method_fig}
\end{figure*}
In this section, we introduce our proposed method, CaDiRec, which is shown in Figure \ref{method_fig}. We commence this section by introducing the Transformer-based SR model. Next, we present the details of our proposed context-aware diffusion-based contrastive learning method. Finally, we introduce the end-to-end training objective of the whole framework.


\subsection{Sequential Recommendation Model}
Similar to many previous studies \cite{qiu2022contrastive,qin2023meta,chen2022intent}, our framework uses a Transformer-based architecture for SR task, which comprises the embedding layer, the transformer layer, and the prediction layer. 
\subsubsection{Embedding Layer} \label{emb layer}
We create an item embedding matrix $ \mathbf{M} \in \mathbb{R}^{|\mathcal{V}| \times d} $ for the item set, where $d$ represents the latent dimensionality. Given a user sequence $\mathbf{s}=[v_1, v_2,...,v_n]$ where $n$ is the max sequence length, we can obtain the input embedding vectors $\mathbf{e}=[\mathbf{e}_1,\mathbf{e}_2,...,\mathbf{e}_n] \in \mathbb{R}^{n\times d}$ with respect to $\mathbf{s}$. 
In addition,  we also construct a position embedding matrix $ \mathbf{P} \in \mathbb{R}^{n \times d} $.
For $w$-th item of the sequence, we add the item embedding $\mathbf{e}_w$ and the corresponding position embedding $\mathbf{p}_w$, resulting the final input vector at step $w$
$\mathbf{h}^0_w = \mathbf{e}_w + \mathbf{p}_w$, and $\mathbf{h}^0=[\mathbf{h}^0_1, \mathbf{h}^0_2,...,\mathbf{h}^0_n]$ denotes the representation of input sequence.

\subsubsection{User Sequence Encoder} \label{user seq enc}
Following the embedding layer, the input vector $\mathbf{h}^0$ is passed through $L$ Transformer blocks to learn the user sequence representations.
Each Transformer (Trm) block consists of a self-attention layer and a feed-forward network layer, which can be formulated as:
\begin{equation}
\label{eq6}
    \mathbf{h}^L = \textbf{Trm}(\mathbf{h}^0 ) ,
\end{equation}
where $\mathbf{h}^L \in \mathbb{R}^{n \times d}$ denotes the hidden states of the last layer, and the vector of the last position $\mathbf{h}^L_n \in \mathbb{R}^{d}$ is used to represent the whole user sequence.

\subsubsection{Prediction Layer}
In the prediction layer, we first calculate the similarities between the user sequence representation vector $\mathbf{h}^L_n$ and item embedding vectors through an inner-product as:
\begin{equation}
    \mathbf{r} = \mathbf{h}^L_n  \mathbf{M}^\mathrm{T},
\end{equation}
where $\mathbf{r} \in \mathbb{R}^{|\mathcal{V}|}$, and $r_i$ is the likelihood of $v_i$ being the next item.
The items are then ranked based on $\mathbf{r}$ to generate the top-k recommendation list.

During training, we adopt the Binary Cross-Entropy (BCE) loss with negative sampling to train the SR model, following many previous methods~\cite{kang2018self,zhou2020s3,xie2022contrastive}.
\begin{equation}
\label{rec_loss}
    \mathcal{L}_{rec} = - \sum_{u\in \mathcal{U}} \sum_{t=1}^n \log\big(\sigma(\mathbf{h}^L_t\cdot \mathbf{e}_{v_{t+1}})\big) + \log\big(1- \sigma(\mathbf{h}^L_t\cdot \mathbf{e}_{v_j^-})\big),
\end{equation}
where we pair each ground-truth item ${v_{t+1}}$ with one negative item ${v_j^-}$that is randomly sampled from the item set.

\subsection{Context-aware Diffusion-based Contrastive Learning}

In this section, we introduce the process of our proposed context-aware diffusion-based contrastive learning method.
Existing methods neglect the context information of user sequences, thereby potentially generating unreasonable augmented views for contrastive learning. In contrast to these methods, we propose to utilize context information as a guidance to generate more reasonable augmented views through conditional generation. Different augmented views of the same user are considered as a pair of positive samples for contrastive learning.

Specifically, given a sequence $\mathbf{s}^u$ for user $u$, we select a subset of items within $\mathbf{s}^u$ based on a predefined ratio $\rho$. The position indices of the selected items within the sequence are recorded as $\mathbf{a}_1^u$. We then employ a context-aware diffusion model to generate items that align with the context information. These generated items replace the original items at the $\mathbf{a}_1^u$ positions, resulting in the augmented sequence $\mathbf{s}^u_1$.
That is, the sole distinction between the original sequence $\mathbf{s}^u$ and the augmented sequence $\mathbf{s}^u_1$ is the replacement of selected items from $\mathbf{s}^u$ with context-aligned items generated by the diffusion model.
The details of the proposed diffusion model will be introduced in Sec. \ref{denoising}. 
By repeating a similar operation, we can obtain another augmented view $\mathbf{s}^u_2$ with respect to another set of selected position indices $\mathbf{a}_2^u$. Note that our method takes into account contextual information and sequential dependencies. Therefore, the generated augmented views do not disrupt the user's interest preferences and interest evolution. Consequently, two augmented views $\mathbf{s}^u_1$ and $\mathbf{s}^u_2$ of user $u$ can be considered as a pair of positive samples, and their representations should be brought closer.

We adopt the standard contrastive loss function to maximize the representation agreement between two different augmented views of the same user sequence and minimize the agreement between the augmented sequences derived from different users.
Specifically, for $\mathbf{s}^u_1$ and $\mathbf{s}^u_2$, we first obtain their input embeddings and then feed them into the user sequence encoder, as defined in Section \ref{user seq enc}. This process generates their representations, $\tilde{\mathbf{h}}^u_1$ and $\tilde{\mathbf{h}}^u_2$, according to Equation (\ref{eq6}). In this way, we can obtain the representations corresponding to the two augmented views for all users. For user $u$, $\tilde{\mathbf{h}}^u_1$ and $\tilde{\mathbf{h}}^u_2$ are regarded as the positive pair, while the remaining $2(N-1)$ augmented representations within the same batch are treated as negative samples ${\mathbf{H}}^-$, where $N$ is the batch size.
Then, we employ the inner product to assess the representation similarity. Finally, we compute the contrastive loss $\mathcal{L}_\textbf{cl}$ as follows:
\begin{equation}
\mathcal{L}_\text{cl}^u = -\log \frac{\exp\big(\text{sim}(\tilde{\mathbf{h}}^u_1, \tilde{\mathbf{h}}^u_2)\big)}{\exp\big(\text{sim}(\tilde{\mathbf{h}}^u_1, \tilde{\mathbf{h}}^u_2)\big)+ \sum_{\tilde{\mathbf{h}}^- \in {\mathbf{H}}^-}\exp\big(\text{sim}(\tilde{\mathbf{h}}^u_1, \tilde{\mathbf{h}}^-)\big) },
\end{equation}
where $\text{sim}(\cdot)$ denotes the inner product of vectors.

 
 \subsection{Context-aware Diffusion Model}
In this section, we present the details of our proposed context-aware diffusion model, which is shown in the lower half of Figure \ref{method_fig}. Our diffusion model generates items for replacing the original ones by utilizing the context information of the selected positions, thereby achieving context-aligned data augmentation. The diffusion process of our model consists of a forward process with partial position noising and a reverse process with context-conditional denoising.



\subsubsection{Forward Process with Partial Position Noising}
In the forward process, we gradually add noise to the selected items of the user sequence.
Specifically, at the start of the forward process, we incorporate a Markov transition from the discrete sequence $\mathbf{s}$ to a continuous vector $\mathbf{z}_0$ using the embedding map, following Diffusion-LM \cite{li2022diffusion}. This transition is parameterized by:
\begin{equation}
    q_\phi(\mathbf{z}_0|\mathbf{s}) = \mathcal{N}(\mathbf{e}, \beta_0 \mathbf{I}),
\end{equation}
where $\mathbf{e}$ denotes the embeddings corresponding to the sequence $\mathbf{s}$, as defined in Section \ref{emb layer}, $\beta_0$ is the variance scale.
This transformation allows us to integrate the discrete sequence into the standard forward process.
At each forward step $q(\mathbf{z}_t|\mathbf{z}_{t-1})$ where $t \geq 1$, we incrementally add Gaussian noise into the hidden states of the previous time step $\mathbf{z}_{t-1}$, to obtain $\mathbf{z}_{t}$, based on Equation (\ref{eq2}).

Unlike other diffusion models, we selectively apply noise to items at chosen positions with a certain ratio $\rho$ instead of the entire sequence, while retaining the items at the remaining positions (i.e., context information).
This approach allows the hidden vectors at the remaining positions and their relative positions to act as the conditional guidance during the reverse phase, enabling our model to utilize context information for controlling item generation.

\subsubsection{Reverse Process with Context-Conditional Denoising.}  \label{denoising}
In the reverse process, context is used to guide the restoration of the conditional distribution at the corresponding positions step by step.
Specifically, we gradually remove noise starting from $\mathbf{z}_T$ and ultimately recover the original data distribution, which is formulated as:
\begin{equation}
    p_{\theta}(\mathbf{z}_{0:T}) =p(\mathbf{z}_{T}) \prod_{t=1}^T p_\theta(\mathbf{z}_{t-1}|\mathbf{z}_t),
\end{equation}
where for each step $t$,
\begin{equation}
    p_\theta(\mathbf{z}_{t-1}|\mathbf{z}_t) = \mathcal{N}\big(\mathbf{z}_{t-1}; \mu_\theta (\mathbf{z}_t, t),\Sigma_\theta(\mathbf{z}_t,t)\big).
\end{equation}
We set $\Sigma_\theta(\mathbf{z}_t,t)$ to untrained time-dependent constants to simplify the computation~\cite{ho2020denoising,li2022diffusion,gong2022diffuseq}.
Following Diffusion-LM \cite{li2022diffusion}, we also incorporate a trainable rounding step $p_\theta(\mathbf{s}|\mathbf{z}_0) = \prod_{i=1}^n p_\theta (v_i|z_i) $ in the reverse process to map the hidden states back to the embedding space, where $p_\theta (v_i|z_i)$ is a softmax distribution. More details about the rounding step can be found in \cite{li2022diffusion}. 

We use a learnable model $f_\theta(\mathbf{z}_t, t)$ to model the reverse process $ p_\theta(\mathbf{z}_{t-1}|\mathbf{z}_t)$. Note that only the hidden vectors corresponding to items selected in the forward process are subjected to the addition of noise.
Therefore, during the reverse process, the hidden vectors of items at the remaining positions (i.e., context information) as well as their position encoding can serve as a condition to guide the generation.
Here, we require a model architecture capable of effectively encoding contextual information to learn the conditional distribution. However, capturing the complex sequential dependencies between items is challenging. The bidirectional Transformer (BERT) \cite{devlin2018bert,vaswani2017attention} provides a promising solution for this task. With its bidirectional self-attention mechanism and positional encoding, BERT can comprehensively understand context from both preceding and succeeding items. Therefore, we employ the bidirectional Transformer encoder as the model architecture for $f_\theta(\mathbf{z}_t, t)$, as illustrated in the bottom-right corner of Figure \ref{method_fig}. The encoder is constructed by stacking $L^\prime$ BERT layers, each comprising a multi-head self-attention layer and a position-wise feed-forward network. At reverse step $t$, the encoder receives $\mathbf{z}_t$ along with the sequence's positional encoding and the diffusion step encoding, subsequently outputting $\mathbf{z}_{t-1}$.


To train the diffusion model, we compute the variational lower bound following previous methods \cite{gong2022diffuseq,ho2020denoising,li2022diffusion}. As we have incorporated the embedding step and rounding step, the variational lower bound loss $\mathcal{L}_{vlb}$ introduced in Eq. (\ref{vlb_loss}) now becomes as follows:
\begin{equation}
\mathcal{L}_{vlb}^\prime =  \mathop{\mathbb E} \limits_{q_\phi(\mathbf{z}_0|\mathbf{s})} \left[  \mathcal{L}_{vlb}(\mathbf{z}_0)
+ \log q_\phi(\mathbf{z}_0|\mathbf{s})
- \log p_\theta(\mathbf{s}|\mathbf{z}_0)
\right].
\end{equation}
Following DiffuSeq \cite{gong2022diffuseq}, this training objective can be further simplified as:
   \begin{equation}
   \label{eq14}
   \begin{aligned}
     \mathcal{L}_{d} &=   
    \sum_{t=2}^T||\mathbf{z}_0 - f_\theta(\mathbf{z}_t,t) ||^2 + 
    || \mathbf{e} -  f_\theta(\mathbf{z}_1,1)||^2 -
     \log p_\theta (\mathbf{s}|\mathbf{z}_0)  \\
    &\rightarrow 
    \sum_{t=2}^T||\tilde{\mathbf{z}}_0 - \tilde{f}_\theta(\mathbf{z}_t,t) ||^2 + 
    || \tilde{\mathbf{e}} -  \tilde{f}_\theta(\mathbf{z}_1,1)||^2 -
     \log p_\theta (\mathbf{s}|\mathbf{z}_0) ,
    \end{aligned}
\end{equation} 
where $\tilde{\mathbf{z}}_0$, $\tilde{f}_\theta$, and $\tilde{\mathbf{e}}$ denote the part of $\mathbf{z}_0$, ${f}_\theta$, and ${\mathbf{e}}$ corresponding to selected positions, respectively. For a detailed derivation process regarding the transformation of this training objective function, please refer to DiffuSeq \cite{gong2022diffuseq}.

Note that while we only calculate the loss with respect to the selected positions in the first term of Equation (\ref{eq14}), 
the reconstruction of the selected items $\tilde{\mathbf{z}}_0$ also takes into account
the remaining items (i.e., context information) of the sequence due to the bidirectional self-attention mechanism.

\subsubsection{Generating Augmented Views}
During contrastive learning, the diffusion model acts as a data generator to generate reasonable augmented views, thereby improving the contrastive learning. Given a user sequence $\mathbf{s}$, we target to generate context-aligned items for arbitrary position indices $\mathbf{\tau}$.
We first randomly sample $\tilde{\mathbf{z}}_T \sim N(0; \mathbf{I}) $ to replace the 
item embeddings ${\mathbf{e}}$ with respect to selected position indices $\mathbf{\tau}$ to obtain $\mathbf{z}_T$. Then, we can iterate the reverse procedure until we reach the initial state $\mathbf{z}_0$. For each step, we adopt the following operations: 1) performing the rounding step (defined in Sec. \ref{denoising}) on $\mathbf{z}_t$ to map it back to item embedding space; 2) replacing the part of recovered $\mathbf{z}_{t-1}$ that does not belong to selected positions $\mathbf{\tau}$ with the original item embeddings, thereby preserving context information. Finally, through the substitution of generated items into the corresponding positions of the original sequence, an augmented sequence is obtained. Performing the same operation twice with different selected positions for the same user results in a pair of positive samples. Note that due to the different initial random noise, the generated items with the same context information will still exhibit a certain level of diversity, which is also important for contrastive learning.

\subsection{End-to-End Training}
As both of the diffusion model and SR model rely on item embeddings, employing separate sets of item embeddings would result in a misalignment between the representation spaces of two models.
To overcome this challenge, we propose to share item embeddings between the two models, and train the full framework in an end-to-end manner. The final objective function is formulated as:
\begin{equation}
    \mathcal{L} =  \mathcal{L}_{{rec}} + \alpha \mathcal{L}_{cl} + \beta \mathcal{L}_{d},
\end{equation}
where $\alpha$ and $\beta$ are hyperparameters that determine the weightings, primarily used to keep the magnitudes of the various loss terms balanced.
$\mathcal{L}_{{rec}}$ and $\mathcal{L}_{cl}$ represent the loss for SR task and the contrastive learning task, respectively. $\mathcal{L}_{d}$ is the loss for the diffusion model. 

Due to the introduction of diffusion model, our model has a longer training time compared to random augmentation-based SR models like CL4SRec \cite{xie2022contrastive}. However, the training time is comparable to that of diffusion-based SR methods like DreamRec \cite{yang2023generate}. Additionally, our data augmentation and contrastive learning are only performed during training. Consequently, our method has a much faster inference time compared to other diffusion-based SR models \cite{yang2023generate} and the inference time is comparable to random augmentation-based SR methods \cite{xie2022contrastive}.

\section{Experiments}
\subsection{Experimental Settings}
\subsubsection{Datasets}

\begin{table}[t] 
\renewcommand\arraystretch{1.0}
	\centering
	\caption{Dataset description.}  
 \setlength\tabcolsep{4.2pt}
 \scalebox{0.9}{
\begin{tabular}{lrrrrr}
\toprule
Datasets & \#Users  & \#Items  & \#Actions & Avg. Length & Density  \\
\midrule
ML-1m &6,040  &3,953  &1,000,209  &165.6  &4.19\%  \\
Beauty &22,363 &12,101  &198,502  &8.8  &0.07\%   \\
Sports &35,598&   18,357 & 296,337 &8.3  &0.05\%    \\
Toys &19,412 &11,924 &167,597 &8.6 &0.07\%   \\
Yelp &30,431 & 20,033 &316,354 &10.4 &0.05\%  \\
\bottomrule
\end{tabular}}
\label{dataset}
\end{table}

\begin{table*}[t]
\renewcommand\arraystretch{1.0}
\centering
\caption{Performance comparison of different methods on five datasets. Bold font indicates the best performance, while underlined values represent the second-best. CaDiRec achieves state-of-the-art results among all baseline models, as confirmed by a paired t-test with a significance level of 0.01.}
\label{tab:methodcomparenew}
\setlength\tabcolsep{2.0pt}
\scalebox{0.9}{
\begin{tabular}{c|l|cccc|ccccc|cc|c|r}
    \toprule
    Dataset&Metric& BPR-MF & Caser & SASRec & BERT4Rec & $\rm {S^3\text{-}Rec}$ & CL4SRec  &CoSeRec & DuoRec & MCLRec & DiffuASR & DreamRec& \textbf{CaDiRec} &Improv. \\
    \midrule
\multirow{4}*{ML-1m}& 
HR@5 &0.0164 &0.0836 &0.1112 &0.0925 &0.1082 &0.1147 &0.1162  &0.1216 &\underline{0.1298} &0.1105 &0.1205 &\textbf{0.1504} &15.9\% \\
     ~&ND@5 &0.0097 &0.0451 &0.0645 &0.0522 &0.0624 &0.0672 &0.0684 &0.0702 &\underline{0.0824} &0.0637 &0.0813 &\textbf{0.1001} &21.5\% \\
     ~&HR@10 &0.0354 &0.1579 &0.1902 &0.1804 &0.1961 &0.1978 &0.1952 &0.1996 &\underline{0.2047} &0.1892 &{0.2006} &\textbf{0.2282} &11.5\% \\
     ~&ND@10 &0.0158 &0.0624 &0.0906 &0.0831 &0.0922 &0.0932 &0.0974 &0.1003 &0.1055 &0.0903 &\underline{0.1077} &\textbf{0.1251} &16.2\% \\
     \midrule
  
  \multirow{4}*{Beauty}&  HR@5 &0.0122 &0.0256 &0.0384 &0.0360 &0.0387 &0.0401 &0.0404 &0.0422 &0.0437 &0.0388 &\underline{0.0440} &\textbf{0.0495} &12.5\% \\
    ~&ND@5 &0.0071 &0.0147 &0.0249 &0.0216 &0.0244 &0.0258 &0.0265  &0.0264 &\underline{0.0278} &0.0251 &0.0274 &\textbf{0.0314} &12.9\% \\
    ~&HR@10 &0.0298 &0.0342 &0.0628 &0.0601 &0.0646 &0.0651 &0.0648  &0.0669 &\underline{0.0689} &0.0633 &0.0687 &\textbf{0.0718} &4.2\% \\
    ~&ND@10 &0.0132 &0.0236 &0.0321 &0.0308 &0.0327 &0.0322 &0.0334  &0.0336 &\underline{0.0357} &0.0316 &0.0352 &\textbf{0.0386} &8.1\% \\
     \midrule

\multirow{4}*{Sports}& HR@5 &0.0095 &0.0154 &0.0225 &0.0217 &0.0173 &0.0221 &0.0245 &0.0232 &\underline{0.0249} &0.0217 &0.0248 &\textbf{0.0276} &10.8\% \\
     ~&ND@5 &0.0062 &0.0124 &0.0142 &0.0143 &0.0112 &0.0129 &0.0159  &0.0154 &\underline{0.0161} &0.0138 &0.0151 &\textbf{0.0183} &13.7\% \\
     ~&HR@10 &0.0193 &0.0261 &0.0339 &0.0359 &0.0311 &\underline{0.0383} &0.0372 &0.0362 &{0.0382} &0.0322 &0.0374 &\textbf{0.0426} &11.2\% \\
     ~&ND@10 &0.0091 &0.0138 &0.0174 &0.0181 &0.0147 &0.0173 &\underline{0.0205}  &0.0189 &{0.0197} &0.0166 &0.0191 &\textbf{0.0233} &13.7\% \\
     \midrule
  
\multirow{4}*{Toys}& HR@5 &0.0102 &0.0169 &0.0453 &0.0461 &0.0443 &0.0468 &0.0474 &0.0459 &0.0491 &0.0448 &\underline{0.0497}&\textbf{0.0522} &5.0\% \\
     ~&ND@5 &0.0061 &0.0106 &0.0306 &0.0311 &0.0294 &0.0317 &0.0323 &0.0322 &\underline{0.0327} &0.0312 &0.0316 &\textbf{0.0356} &8.9\% \\
     ~&HR@10 &0.0135 &0.0271 &0.0675 &0.0665 &0.0693 &0.0684 &0.0695  &0.0681 &\underline{0.0702} &0.0667 &0.0643 &\textbf{0.0785} &11.8\% \\
     ~&ND@10 &0.0094 &0.0140 &0.0374 &0.0368 &0.0375 &0.0388 &0.0401  &0.0385 &\underline{0.0412} &0.0382 &0.0402 &\textbf{0.0441} &7.0\% \\
 \midrule
 
\multirow{4}*{Yelp}& HR@5 &0.0127 &0.0151 &0.0161 &0.0186 &0.0199 &0.0201 &0.0198  &0.0199 &\underline{0.0209} &0.0157 &0.0174 &\textbf{0.0238} &13.9\% \\
     ~& ND@5 &0.0074 &0.0096 &0.0100 &0.0118 &0.0118 &0.0124 &0.0120 &0.0123 &\underline{0.0129} &0.0102 &0.0116 &\textbf{0.0149} &15.5\% \\
     ~& HR@10 &0.0273 &0.0253 &0.0274 &0.0338 &0.0291 &0.0349 &0.0323  &0.0342 &\underline{0.0354} &0.0268 &0.0245 &\textbf{0.0387} &9.3\% \\
     ~& ND@10 &0.0121 &0.0129 &0.0136 &0.0171 &0.0168 &0.0181 &0.0179  &\underline{0.0189} &0.0177 &0.0133 &0.0152 &\textbf{0.0197} &4.2\% \\
     
    \bottomrule
\end{tabular} }
\label{comparison_new}
\end{table*}

We conduct experiments on five real-world public datasets, including MovieLens, Beauty, Sports, Toys, and Yelp. The statistics of these datasets are shown in Table \ref{dataset}. These datasets encompass a wide range of application scenarios. 
The MovieLens\footnote{https://grouplens.org/datasets/movielens/} dataset is a stable benchmark dataset which collects movie ratings provided by users.
Beauty, Sports, and Toys datasets are obtained from Amazon\footnote{http://jmcauley.ucsd.edu/data/amazon/}, one of the largest e-commerce platforms globally. 
Yelp is a renowned dataset primarily used for business recommendation.
We adopt the same preprocessing method as employed in numerous previous studies \cite{liu2021contrastive,xie2022contrastive}, filtering items and users with fewer than five interaction records. 

\subsubsection{Evaluation Metrics}
To evaluate the performance of our
model and baseline models, we employ widely recognized evaluation metrics: Hit Rate (HR) and Normalized Discounted Cumulative Gain (NDCG), and report values of HR@k and NDCG@k for k=5 and 10. We use the standard leave-one-out strategy, utilizing the last and second-to-last interactions for testing and validation, respectively, while the remaining interactions serve as training data. To ensure unbiased evaluation, we rank all items in the item set and compute the metrics based on the rankings across the entire item set.

\subsubsection{Baseline Methods}
To ensure a comprehensive assessment, we compare our method with eleven baseline methods, which can be divided into three categories: classical methods (BPR-MF, Caser, SASRec, BERT4Rec), contrastive learning-based methods ($\rm {{S}^{3}\text{-}{Rec}}$, CL4SRec, CoSeRec, DuoRec, MCLRec), and diffusion-based methods (DiffuASR, DreamRec).

\begin{itemize}[leftmargin=1em]
\item \textbf{BPR-MF} \cite{rendle2012bpr}. It employs matrix factorization to model users and items, and uses the pairwise Bayesian Personalized Ranking (BPR) loss to optimize the model.
\item \textbf{SASRec} \cite{kang2018self}. It is the first work to utilize the self-attention mechanism for sequential recommendation.
\item \textbf{Caser} \cite{tang2018personalized}. It utilizes a CNN-based approach to model high-order relationships in the context of sequential recommendation.
\item \textbf{BERT4Rec} \cite{sun2019bert4rec}. It employs the BERT \cite{devlin2018bert} framework to capture the context information of user behaviors.
\item $\rm {\textbf{S}^{3}\text{-}\textbf{Rec}}$ \cite{zhou2020s3}. It leverages self-supervised learning to uncover the inherent correlations within the data. However, its primary emphasis lies in integrating the user behavior sequence and corresponding attribute information.
\item \textbf{CL4SRec} \cite{xie2022contrastive}. It proposes three random augmentation operators to generate positive samples for contrastive learning.
\item $\textbf{CoSeRec}$ \cite{liu2021contrastive}. It introduces two informative augmentation operators leveraging item correlations based on CL4SRec. We compare with these informative augmentations.
\item \textbf{DuoRec} \cite{qiu2022contrastive}. It combines a model-level augmentation and a novel sampling strategy for choosing hard positive samples.
\item \textbf{MCLRec} \cite{qin2023meta}. It integrates both data-level and model-level augmentation strategies, utilizing CL4SRec's random data augmentation for the input sequence and employing MLP layers for model-level augmentation.
\item \textbf{DiffuASR} \cite{liu2023diffusion}. It leverages the diffusion model to generate pseudo items and concatenates them at the beginning of raw sequences. Then, the extended sequences are fed into a downstream recommendation model for next item prediction.
\item \textbf{DreamRec} \cite{yang2023generate}. It directly utilizes the diffusion model to generate the next item based on the historical interactions.
\end{itemize}

\subsubsection{Implementation Details} We implement all baseline methods according to their released code. The embedding size for all methods is set to 64. Our method utilizes a Transformer architecture for the SR model, comprising 2 layers and 2 attention heads each layer. Meanwhile, our diffusion model employs a bidirectional Transformer with 1 layer and 2 attention heads. The total number of diffusion steps is set to a fixed value of 1000. 
We tune the coefficients of the two critical terms in the loss function, $\alpha$ and $\beta$ within the range of [0.1, 0.2, 0.4, 0.6, 0.8, 1.0]. Additionally, we explore the substitution ratio $\rho$ within the range of [0, 0.1, 0.2, 0.4, 0.6, 0.8].
The Dropout rate is chosen from the set \{0.1, 0.2, 0.3, 0.4, 0.5\} for both the embedding layer and the hidden layers. We set the training batch size to 256 and employ the Adam optimizer with a learning rate of 0.001.
Following most previous works \cite{kang2018self}, we set the max sequence length to 50 for three Amazon datasets and Yelp, and to 200 for the MovieLens dataset.

For the SR task, the majority of baseline models employ the negative sampling strategy during the training process, while others calculate the probability across the entire item set using the softmax function, which is impractical when dealing with considerably large item sets.
In our initial experiments, we observed that the two distinct training strategies significantly impact the results. To facilitate a fair comparison focused solely on assessing the impact of contrastive learning, we employ the BCE loss with the negative sampling strategy (defined in Equation (\ref{rec_loss})) for all methods.




\begin{table}[t]
\renewcommand\arraystretch{1.0}
\centering
\caption{Ablation study on five datasets.}
\setlength\tabcolsep{4.0pt}
\scalebox{0.85}{
\begin{tabular}{l|c|ccccc}
\toprule
 & Metric & {\makecell[c]{w/o CG}} & {\makecell[c]{w/o B-Enc}} & {\makecell[c]{w/o $\mathcal{L}_{d}$}} &{\makecell[c]{w/o  $\mathcal{L}_{cl}$}} & CaDiRec  \\
\midrule
\multirow{2}*{ML-1m}  &HR@10  & 0.1762 & 0.2203 & 0.1757 & 0.1932 & 0.2282 \\
  &ND@10  &0.0861  &0.1212  &0.0855  &0.0982  &0.1251  \\
  \midrule
\multirow{2}*{Beauty} &HR@10  & 0.0647 & 0.0695 & 0.0644 & 0.0673 & 0.0718 \\
  &ND@10  &0.0355  &0.0365  &0.0353  &0.0359  &0.0386  \\
  \midrule
\multirow{2}*{Sports} &HR@10  & 0.0361 & 0.0399 & 0.0363 & 0.0391 & 0.0426 \\
  &ND@10  &0.0202  &0.0211  &0.0199  &0.0208  &0.0233  \\
  \midrule
\multirow{2}*{Toys} &HR@10  & 0.0698 & 0.0738 & 0.0695 & 0.0721 & 0.0785 \\
  &ND@10  &0.0398  &0.0419  &0.0396  &0.0412  &0.0441  \\
  \midrule
\multirow{2}*{Yelp} &HR@10  & 0.0304 & 0.0351 & 0.0300 & 0.0312 & 0.0387 \\
  &ND@10  &0.0149  &0.0172  &0.0148  &0.0153  &0.0197  \\
\bottomrule
\end{tabular}}
\label{ablation study}
\vspace{-1em}
\end{table}

\subsection{Comparison Results}
We run each experiment five times and report the average results. The comparison results across all datasets are presented in Table \ref{tab:methodcomparenew}. 
Based on these results, we make the following observations:
\begin{itemize}[leftmargin=1em]
\item Our method consistently outperforms all eleven baseline models across all datasets. Additionally, a paired t-test reveals that our method achieves significantly better performance than the second-best result, with a significance level of 0.01.
\item Classical methods (BPR-MF, Caser, SASRec, BERT4Rec) that do not employ contrastive learning tend to perform poorly compared to methods that integrate data augmentation and contrastive learning. This suggests that contrastive learning, serving as an auxiliary task, facilitates more comprehensive learning of user sequence representations in the presence of limited data, thereby improving sequential recommendation.
\item Our method consistently outperforms contrastive learning-based baselines ($\rm{{S}^{3}\text{-}{Rec}}$, CL4SRec, CoSeRec, DuoRec, MCLRec) across all metrics on all datasets. CL4SRec introduces three random data augmentation operations for contrastive learning based on SASRec, achieving better performance. CoSeRec takes into account item similarity based on random augmentation, outperforming CL4SRec. DuoRec and MCLRec further improve contrastive learning-based sequential recommendation by incorporating model-level learnable augmentation, resulting in certain improvements. However, all these baseline models neglect context information during augmentation, which may lead to unreasonable positive pairs. Our model, in contrast, leverages context information to guide the generation of augmented views, resulting in superior performance.

\item Our model performs significantly better than existing diffusion-based methods (DiffuASR, DreamRec). DiffuASR does not perform well, likely because its augmentation strategy resembles the reverse multi-step sequential recommendation task, which is extremely challenging and prone to introducing noisy data. Furthermore, DiffuASR feeds these extended sequences to the recommendation model, which may lead to error accumulation. DreamRec, on the other hand, directly uses the diffusion model to generate the next item based on historical items. Unlike these two diffusion-based baselines, CaDiRec uses the diffusion model to generate more reasonable augmented user sequences for better contrastive learning. With context guidance, CaDiRec generates alternative items that adhere to the learned context-conditional distribution. The results show that CaDiRec consistently outperforms both diffusion-based baselines across all datasets.

\end{itemize}

\vspace{-0.5em}
\subsection{Ablation Study}
In this section, we demonstrate the effectiveness of our model by comparing its performance with five different versions across five datasets. The results are shown in Table \ref{ablation study}, where ``w/o CG'' denotes removing context guidance, ``w/o B-Enc'' denotes removing the BERT encoder (utilizing an MLP encoder instead), ``w/o $\mathcal{L}_{d}$'' means removing the diffusion loss term, and ``w/o $\mathcal{L}_{cl}$'' means removing the contrastive learning loss term.
Specifically, when context information is removed, the model's performance significantly decreases across all datasets, highlighting the substantial contribution of context information. The context guidance allows the model to generate more reasonable augmented views, thus enhancing the quality of contrastive learning. When using an MLP encoder instead of the BERT encoder to model context information, performance also declines, indicating that the BERT encoder is more effective at capturing contextual dependencies, thereby providing better guidance for data augmentation. Furthermore, removing $\mathcal{L}_{d}$ results in a performance drop because the diffusion model is not involved in the training update, equivalent to random augmentation, which can lead to unreasonable augmented positive sample pairs. Finally, the decline in performance upon removing $\mathcal{L}_{cl}$ underscores the importance of the contrastive learning task, which has been validated in many previous studies \cite{qiu2022contrastive}. Overall, the results indicate that removing any component reduces the model's performance, thereby validating the effectiveness of each module.


\begin{figure}[t]
\setlength{\abovecaptionskip}{-0.1mm} 
\setlength{\belowcaptionskip}{-1mm} 
  \centering
  \includegraphics[width=0.85\columnwidth]{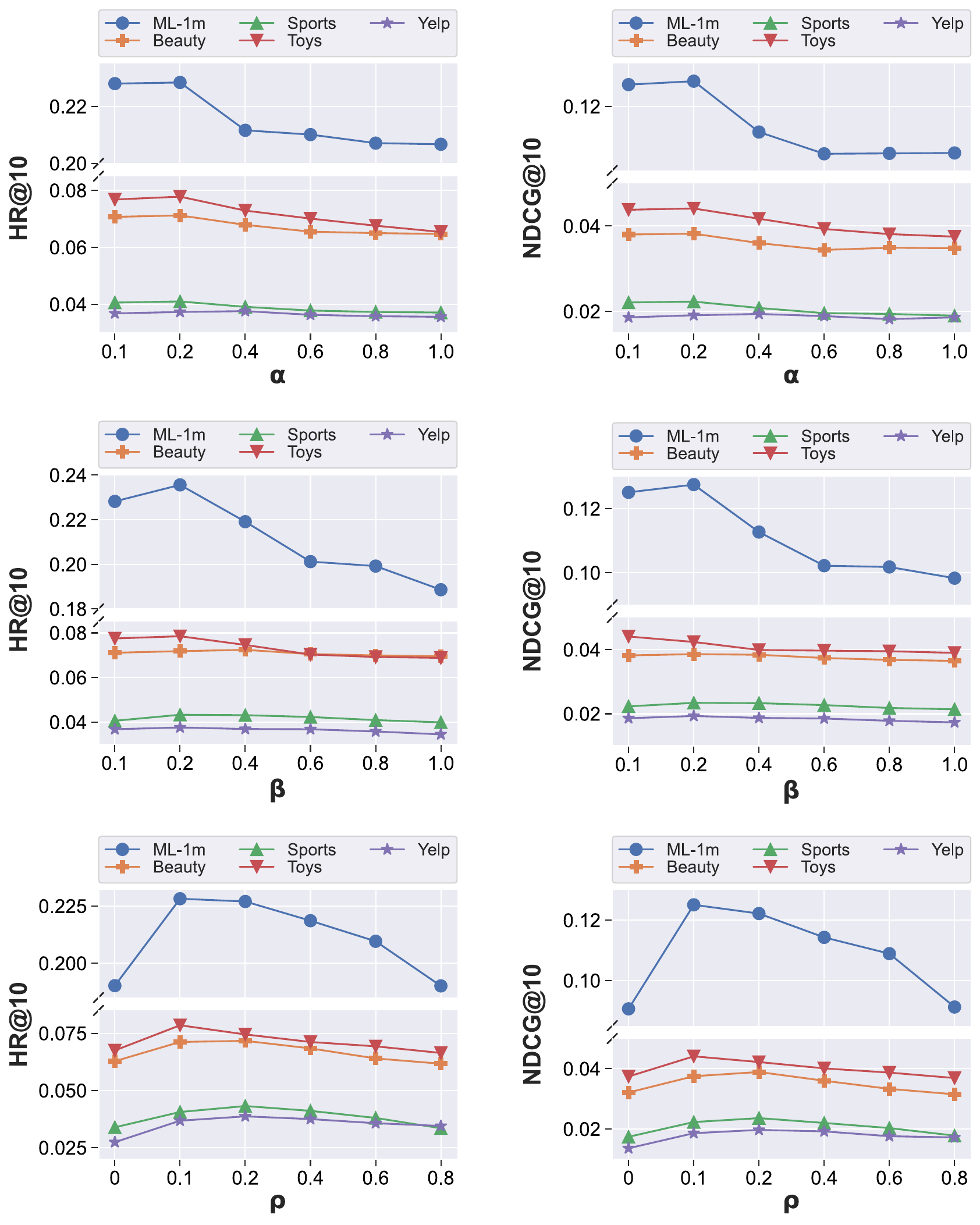}
  \caption{Hyperparameter study of $\alpha$, $\beta$, and $\rho$ on five datasets.} 
   \label{param_study}
\end{figure}

\subsection{Hyperparameter Study}
In this section, we investigate the impacts of three important hyperparameters ($\alpha$, $\beta$, and $\rho$) on HR@10 and NDCG@10 across all five datasets. Here, $\alpha$ represents the weight of the contrastive learning loss, $\beta$ is the weight of the diffusion loss, and $\rho$ is the substitution ratio. The results are shown in Figure \ref{param_study}. 
We observe that as $\alpha$ increases, HR@10 and NDCG@10 initially rise slightly and then decline across all datasets, with the optimal value at approximately $\alpha = 0.2$. $\beta$ controls the weight of the diffusion loss in the total loss. As $\beta$ varies, HR@10 and NDCG@10 values show minimal changes, with an overall trend of initial increase followed by a slight decline. The model achieves optimal performance on all datasets with $\beta \leq 0.4$. 
As the substitution ratio $\rho$ gradually increases from 0 to 0.8 (note that $\rho=1$ represents removing context, as shown in the ablation study), the model's performance initially improves and then declines. The optimal performance is observed when $\rho$ is approximately 0.1 to 0.2. This can be explained by the reduction of context information as $\rho$ increases; without adequate context guidance, the model is unable to generate reasonable positive samples.
When $\rho=0$, no replacements are made, which is equivalent to not using contrastive learning, resulting in poor performance. Therefore, to enhance the effectiveness of contrastive learning, it is advisable to select an appropriate $\rho$ for data augmentation. 
Additionally, the metrics for the MovieLens dataset vary differently with changes in hyperparameters compared to the other four datasets. This difference is due to the fact that the other four datasets are sparse, while MovieLens is relatively dense.

\begin{figure}[t]
\setlength{\belowcaptionskip}{-5mm} 
  \centering
  \includegraphics[width=0.9\columnwidth]{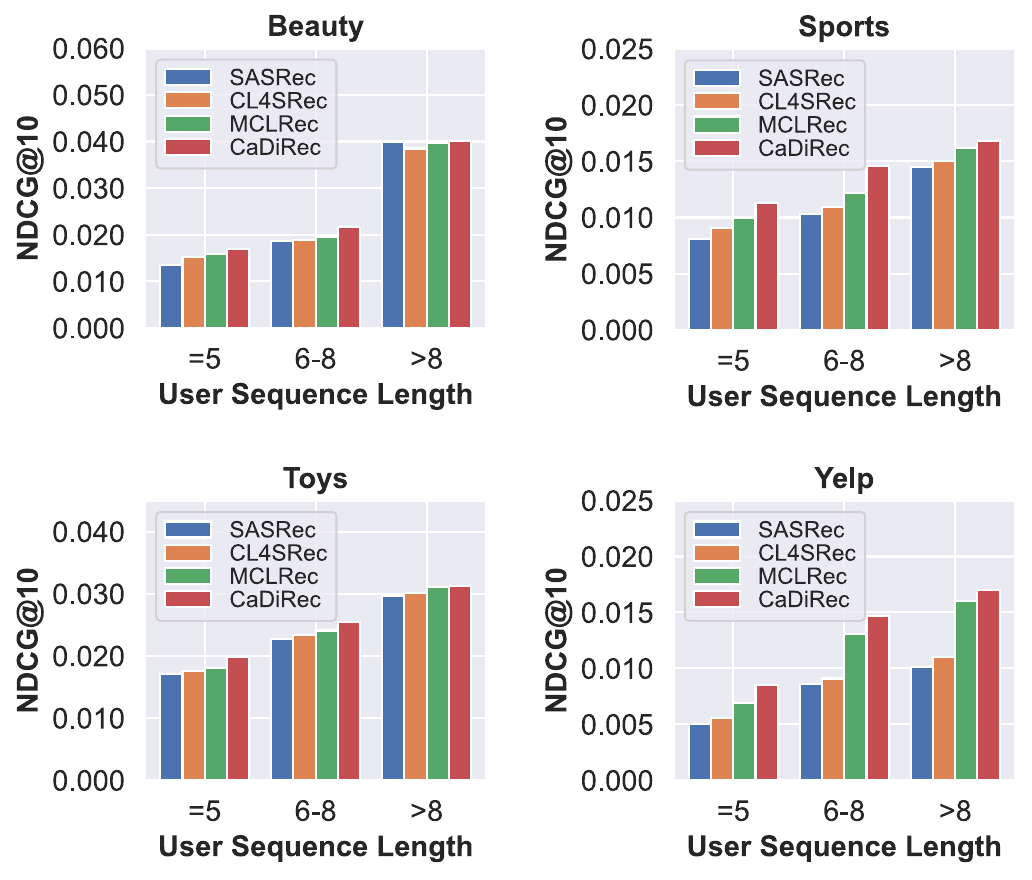}
  \caption{Performance comparison on different user groups.} 
   \label{group study}
\end{figure}

\subsection{Robustness w.r.t. User Sequence Length}
To further examine the robustness of our model against varying degrees of data sparsity, particularly its performance with limited interaction records, we categorize user sequences into three groups based on their length and analyze the evaluation results for each group. Figure \ref{group study} presents the comparison results on the four sparse datasets (excluding MovieLens, as it is a dense dataset).
By comparing our model with representative baseline models, including the strongest baseline MCLRec, we make the following observations:
1) The performance of all models deteriorates as interaction frequency decreases, indicating the influence of data sparsity on model performance.
2) Our model consistently outperforms the baseline models in each user group. Even for the group with the most limited data (sequence length of 5), our model maintains a significant lead, demonstrating the positive impact of our context-aware diffusion-based contrastive learning approach in addressing data sparsity.
This finding underscores the robustness of our model across various degrees of data sparsity in user sequences.

\subsection{Sequence Representation Visualization}

\begin{figure}[t]
\setlength{\belowcaptionskip}{-5mm} 
  \centering
  \includegraphics[width=0.8\columnwidth]{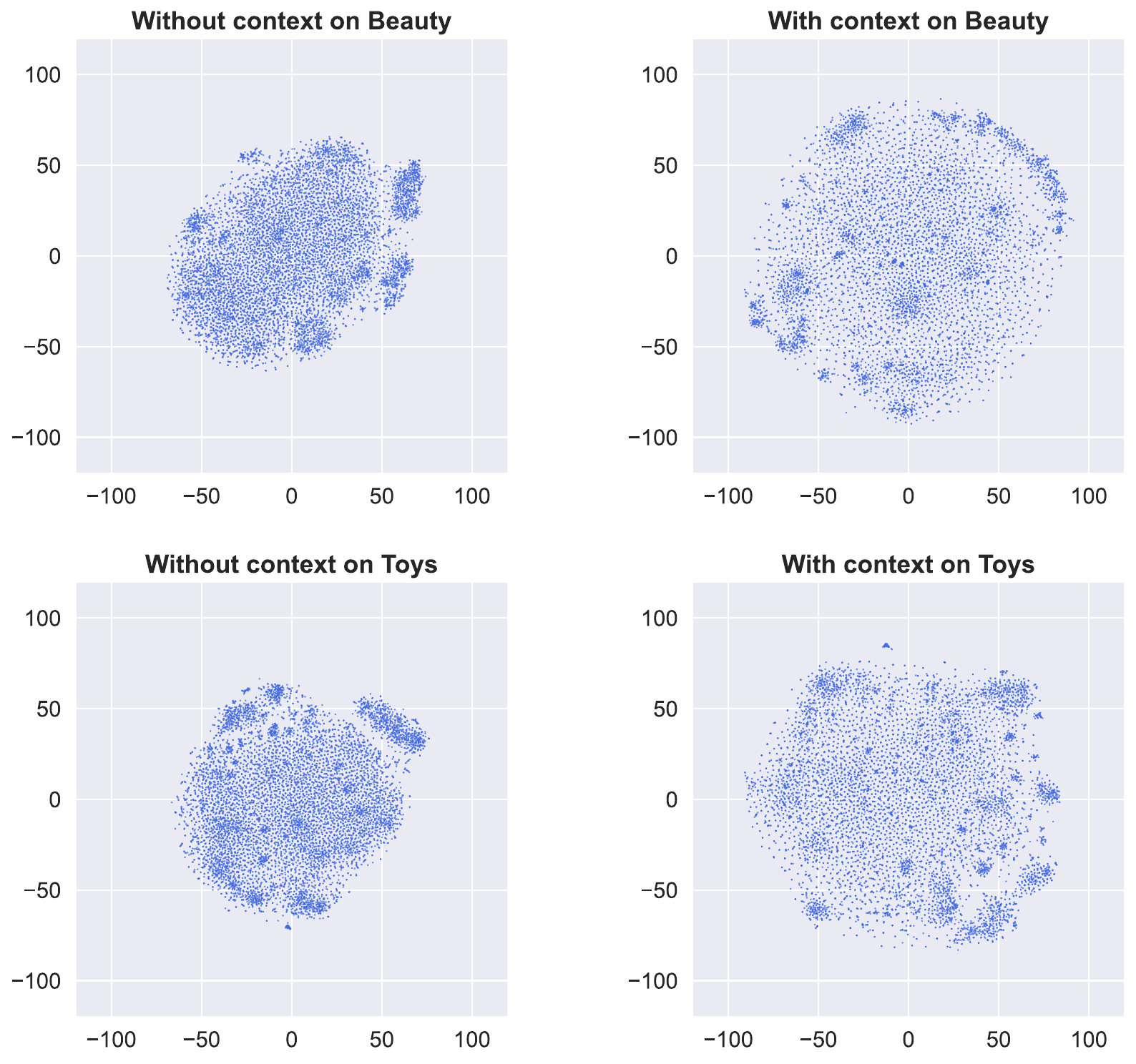}
  \caption{Visualization of learned sequence representations.} 
   \label{visualization}
\end{figure}

To further analyze the impact of context information on representation learning, we visualize the user sequence representations learned by our model with and without context guidance. For both versions of the model, we train it for 300 epochs in an end-to-end manner and utilize t-SNE \cite{van2014accelerating} to reduce the learned user representations to two-dimensional space. Due to space limitations, the results for Beauty and Toys are presented in Figure \ref{visualization}.
Intuitively, the embeddings with and without context exhibit different levels of dispersion in the visualizations. The embeddings without context appear overly compact, while the embeddings with context are comparatively more dispersed, suggesting richer and more informative representations. This pattern is consistent across both datasets.
This may be because augmentations without context resemble random augmentations, which easily generate unreasonable positive sample pairs. In such cases, the contrastive learning objective forcibly brings together dissimilar user sequences, leading to overly compact user representations in the embedding space and even a tendency for representation collapse. Conversely, using context information to guide the generation of augmented views results in more reasonable augmentations, effectively addressing this issue and preventing representation collapse.

\section{Conclusion}
In this paper, we propose a context-aware diffusion-based contrastive learning method for sequential recommendation. We employ a diffusion model to generate more reasonable augmented sequences through conditional generation, thereby improving contrastive learning. We conduct extensive experiments and analyses on five public benchmark datasets. The results demonstrate the advantages of our proposed method over existing baselines.

\section{Acknowledgments}
This work was supported by the Start-up Grant (No. 9610564), the Donations for Research Projects (No. 9229129) of the City University of Hong Kong, and the Early Career Scheme (No. CityU 21219323) of the University Grants Committee (UGC).

\bibliographystyle{ACM-Reference-Format}
\bibliography{reference}

\end{document}